\documentclass[tightenlines,showpacs,prl, floatfix,preprintnumbers,amsmath,amssymb,amsbsy,twocolumn]{revtex4}

\usepackage{graphicx}

\usepackage{bm}

\begin{document}
\vspace{1cm}

\title{Accessibility of the Gravitational-Wave Background due to Binary Coalescences to Second and Third Generation Gravitational-Wave Detectors}

\author{C. Wu$^a$, V. Mandic$^a$ and T. Regimbau$^b$}
\affiliation{$^a$School of Physics and Astronomy, University of Minnesota, Minneapolis, MN 55455, USA\\
$^b$Departement Artemis,  Observatoire de la C\^ote d'Azur, CNRS, F-06304 Nice,  France}

\date{\today}

\begin{abstract}
Compact binary coalescences, such as binary neutron stars or black holes, are among the most promising candidate sources for the current and future terrestrial gravitational-wave detectors. While such sources are best searched using matched template techniques and chirp template banks, integrating chirp signals from binaries over the entire Universe also leads to a gravitational-wave background (GWB). In this paper we systematically scan the parameter space for the binary coalescence GWB models, taking into account uncertainties in the star formation rate and in the delay time between the formation and coalescence of the binary, and we compare the computed GWB to the sensitivities of the second and third generation gravitational-wave detector networks. We find that second generation detectors are likely to detect the binary coalescence GWB, while the third generation detectors will probe most of the available parameter space. The binary coalescence GWB will, in fact, be a foreground for the third-generation detectors, potentially masking the GWB background due to cosmological sources. Accessing the cosmological GWB with third generation detectors will therefore require identification and subtraction of all inspiral signals from all binaries in the detectors' frequency band.
\end{abstract}


\bibliographystyle{plain}
\maketitle

\pagestyle{plain}
\section{1. Introduction}
The ground-based gravitational-wave detectors are rapidly increasing their sensitivities. The first generation detectors LIGO
\cite{LIGOinstr,S5detector} and Virgo \cite{Virgo1,Virgo2} have
reached their design sensitivities and collected excellent data over several years of exposure. The second generation detectors, Advanced LIGO \cite{aLIGO, aLIGO2}, Advanced Virgo \cite{aVirgo}, GEO-HF \cite{GEOHF}, and LCGT \cite{CLIO,LCGT} are currently being built and commissioned. With 10 times better strain sensitivity, these detectors are expected to yield first direct detections of gravitational-wave signals, and their first data is expected as early as 2014. Furthermore, there are already efforts under way to design the third-generation gravitational wave detectors, with another factor of 10 improvement in sensitivity. This includes the Einstein Telescope project \cite{ET,ET2}, for which the design study was recently completed. These detectors are expected to open a new era in astronomy and astrophysics, providing new observations of various events and objects in the Universe, complementary to the standard electromagnetic observations.

Among the many sources of gravitational waves, the coalescences of binary systems, such as binary neutron stars (BNS), binary black holes (BBH), or a black hole and a neutron star (BHNS) stand out as the most likely candidates for first detections. These systems generate well understood "chirp" gravitational-wave signals, which have been computed using post-Newtonian approximation \cite{phenom_waveform} or numerical relativity simulations \cite{numrel}. One can then search for the chirp signals using matched template techniques - indeed a number of such searches have been performed using LIGO and Virgo data \cite{CBC1,CBC2,CBC3}.

It has also been argued that adding the gravitational-wave signals from all binaries in the Universe will produce a gravitational-wave background (GWB) - for example, see \cite{regman,regfrei,regchauv,regreview,ros11,mar11,BBHstoch} for the most recent studies in the context of terrestrial detectors. The LIGO and Virgo collaborations have developed techniques for searching for GWB by cross-correlating data from pairs of gravitational wave detectors \cite{allenromano}. Such searches have also been performed using LIGO and Virgo data \cite{S3stoch,S4stoch,S5stoch}, and have produced competitive upper limits on the energy density carried by gravitational waves.

The goal of this paper is to perform a detailed study of the accessibility of the GWB produced by the binary coalescences to the second and third generation gravitational-wave detectors. Our study follows the work of Regimbau and Mandic \cite{regman}, and includes detailed scans of the parameter space in these models, as well as possible effects due to the uncertainty in the star formation rate and in time-delays associated with the formation of the binaries.
We will show that this background is likely to be observed by the network of second generation detectors, and that the third-generation detectors will likely be able to explore most of the parameter space for these models. In Section 2 we summarize the calculation of the energy density for these models. In Section 3 we present results of our systematic study, and we include concluding remarks in Section 4.

\section{2. Calculation of the Energy Spectrum}
The energy spectrum of gravitational waves is usually described by the dimensionless parameter:
\begin{eqnarray}
\Omega_{gw}(f) & = & \frac{f}{\rho_c} \frac{d\rho_{\rm GW}(f)}{df}
\end{eqnarray}
where $f$ is frequency, $d\rho_{\rm GW}/df$ is the energy density in the frequency range $[f, f+df]$ and
$\rho _{c}$ is the critical energy density needed to close the Universe:
\begin{equation}
\rho _{c}=\frac{3H_{0}^{2}c^2}{8\pi G}.
\end{equation}
where $H_0$ and $G$ are the Hubble parameter and Newton's constant respectively and $c$ is the speed of light.

The energy spectrum for the case of binary coalescences can be written as follows (see for instance \cite{regman, regreview,ros11,mar11}, we will follow \cite{regman} in our approach):
\begin{equation}
\Omega_{\rm GW}(f)=\frac{1}{\rho_c c} f F(f)
\end{equation}
where the integrated flux (per unit frequency) is defined as:
\begin{equation}
F(f)= \int  R_z(z) \frac{1}{4 \pi d^2_L(z)} \frac{dE_{\rm{GW}}(f)}{df}dz
\label{eq-flux}
\end{equation}
where $R_z(z)$ is the rate of gravitational-wave sources per interval of redshift $z$ as observed in the detector (Earth) frame, $d_L(z)=(1+z) r(z)$ is the luminosity distance, $r(z)$ is the proper distance, and $\frac{dE_{\rm{GW}}}{df}$ is the gravitational spectral energy emitted by a single source and observed in detector frame.
The rate in Eq. \ref{eq-flux} is given by:
\begin{equation}
R_z(z)= \lambda R_V(z) \frac{dV(z)}{dz}
\end{equation}
where $\lambda$ is the mass fraction converted into progenitors (discussed in more detail below), $R_V(z)$ is the observed rate of binary coalescences (in units of mass per unit comoving volume per time), and
\begin{equation}
\frac{dV(z)}{dz}= \frac{4 \pi c}{H_0} \frac{r^2(z)}{E(\Omega_{\rm M},\Omega_{\Lambda},z)}
\end{equation}
with $E(\Omega_{\rm M},\Omega_{\Lambda},z) = \sqrt{\Omega_{\rm M} (1+z)^3 + \Omega_{\Lambda}}$ capturing the dependence of the comoving volume on redshift. We use the standard $\Lambda$CDM cosmology, with $\Omega_{\rm M} = 0.3$,
$\Omega_{\Lambda} = 0.7$, and Hubble parameter $H_0 = 70 {\rm \; km \; s^{-1} \; Mpc^{-1}}$.
The rate $R_V(z)$ is dependent on both the star formation rate and on the time-delay $t_d$ between the formation of the binary system and the actual coalescence, and can be written in the following form:
\begin{equation}
R_V(z) = \int \frac{1}{1+z_f} \; R_*(t_c(z) - t_d) P(t_d) dt_d
\end{equation}
where $R_*$ is the star formation rate (discussed further below), $t_c(z)$ is the cosmic time corresponding to redshift $z$, and $z_f$ is the redshift at the formation time $t_c(z) - t_d$. The factor $(1+z_f)$ in the denominator corrects for the time dilation due to the cosmic expansion and converts the rate from the source frame into the detector frame.

Population synthesis \cite{popsynth,dominik} suggests that the probability distribution
for the delay time is well described by $P(t_d) \sim t_d^{\alpha}$ for $t_d>t_{\min}$, where $t_{\min}$ is the minimum delay time for a massive binary to evolve until coalescence. While the currently preferred parameter values are $\alpha = -1$ and $t_{\min}=20$ Myr for BNS and 100 Myr for BBH, other values cannot be excluded. Following \cite{ando}, we will investigate the following ranges for these parameters: $\alpha = -0.5, -1, -1.5$, $t_{\min} = 20, 100$ Myr for BNS, and $t_{\min} = 100, 500$ Myr for BBH. We will also examine the case where time delay is ignored, $t_d=0$.
The star formation rate $R_*$ has also been investigated by several authors \cite{hopkins,fardal,wilkins,nagamine,springel}, for a recent review see \cite{regreview,reghugh}. We will investigate dependence of our results on the choice of the star-formation rate.

The scaling factor $\lambda$ (in units of M$^{-1}_{\odot}$) is a parameter  which includes three different effects: for BNS, these are the mass fraction of neutron star progenitors, the fraction of massive binaries formed among all stars, and the fraction of binaries that remain bounded after the second supernova event (and similarly for the BBH and BHNS). All of these factors are associated with significant uncertainties, which is why we will treat $\lambda$ as a free parameter of the model in our study. We note, however, that $\lambda R_V(0)$ represents the local (present) rate of binary coalescences. These local rates have been a subject of multiple studies, as they directly impact the number of individual binary coalescences that could be detected by the second-generation gravitational-wave detectors. A recent study by the LIGO Scientific Collaboration has produced pessimistic, realistic, optimistic, and maximal possible estimates for these rates \cite{LIGOrates}, based on the observed galactic binary pulsars and on the population-synthesis models. We will compare the results of our study to these rates estimates.

The final factor appearing in the Equation \ref{eq-flux} is $dE_{\rm GW}/d f$, the gravitational spectral energy from a single source. For the BNS and BHNS models, we will only include the inspiral part of the gravitational-wave signal. In the quadrupolar approximation, and assuming circular orbit,  the observed spectral GW energy, averaged over orientation, from a binary system at redshift $z$ is given by a rather simple form:
\begin{equation}
\frac{dE_{\rm GW}}{df} =  \frac{(G\pi)^{2/3}}{3}  (M_c^z)^{5/3} f^{-1/3}
\end{equation}
where $M_c^z=(1+z) M_c$ is the observed redshifted  chirp mass of the binary system and  $M_c$ is the physical mass. We will assume the following ranges for $M_c$: 1-2.5 $M_{\odot}$ for BNS, 2.5-10 $M_{\odot}$ for the BHNS, and 2.5-20 $M_{\odot}$ for the BBH models. These mass ranges are based on \cite{dominik}, allowing for uncertainties in possible neutron star and black hole masses. For the BBH case, however, we will follow
\cite{mar11,BBHstoch} and use the more complex functional form derived by \cite{BBHstoch}, which includes the inspiral, merger, and ringdown contributions to the gravitational-wave signal.

The upper limit on the integral range in Eq. \ref{eq-flux} depends on both the emission frequency range, $f_{\min} - f_{\max}$, in the source frame, and on the maximum redshift $z_{\max}$ considered for the star formation history calculation:
\begin{equation}
z_{\sup} (f)= \left\lbrace
\begin{array}{ll}
z_{\max}    &   \hbox{  if } f < \frac{f_{\max} }{(1+z_{\max})}\\
\frac{f_{\max}}{f}-1 &   \hbox{  otherwise }\\
\end{array}
\right.
\label{eq-zsup}
\end{equation}

Combining the expressions above, we obtain for the density parameter:
\begin{equation}
\Omega_{gw}(f)=\frac{8\lambda (\pi G M_c)^{5/3}}{9 H_0^3 c^2}  f^{2/3}
\int^{z_{\sup}}_{z_{\inf}}  \frac{R_V(z) dz}{(1+z)^{1/3} E(\Omega_{\rm M},\Omega_{\Lambda},z)}
\label{omega}
\end{equation}
Unless noted otherwise, we set $z_{\inf} = 0$ in our calculation.

\begin{figure}[h]
\includegraphics[width=3in]{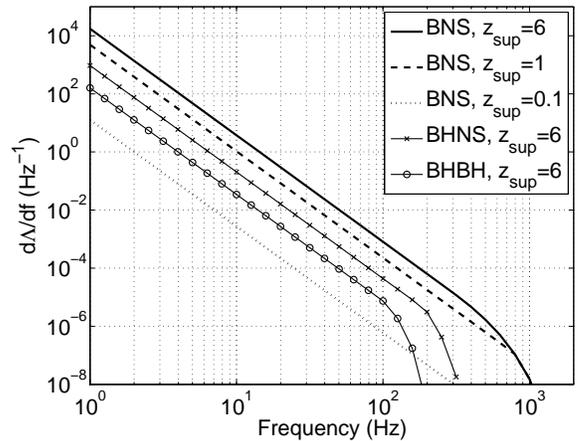}
\caption{Number of binaries per 1 Hz frequency bin.
For the BNS cases we assume each star to have mass of 1.4 $M_{\odot}$, local rate of $\lambda = 1 {\rm \; Mpc^{-3} \; Myr^{-1}}$, and $t_{\min} = 20$ Myr ($P(t_d) \sim 1/t$). For the BHNS case we assume masses of 1.4 $M_{\odot}$ and 10 $M_{\odot}$, the local rate $\lambda = 0.03 {\rm \; Mpc^{-3} \; Myr^{-1}}$, and $t_{\min} = 100$ Myr. For the BBH case we assume masses of 10 $M_{\odot}$, and $t_{\min} = 100$ Myr.}
\label{Lambda}
\end{figure}

We emphasize, however, that the GWB computed here is not necessarily continuous in time, as already noted in \cite{ros11,BBHstoch,reghugh,cowreg}. To illustrate this we compute a duty cycle parameter, defined as
\begin{equation}
\frac{d\Lambda}{df} = \int_0^{z_{\sup}} R_z(z) \frac{d\tau(z)}{df} dz
\end{equation}
where $d\tau(z)/df$ represents the time a binary spends in the frequency band $[f,f+df]$ after properly accounting for redshift:
\begin{equation}
\frac{d\tau}{df}  = \frac{5 c^5}{96 \pi^{8/3}G^{5/3}} (M^z_c)^{-5/3} f^{-11/3}.
\end{equation}
The quantity $d\Lambda/df$ then captures the number of binaries generating gravitational-wave signals in a 1 Hz bin as observed by a detector on Earth (it can be compared to the overlap function of \cite{ros11}, which is the number of sources present on average in a frequency bin $\Delta f$ around the frequency $f$).
This quantity is plotted in Figure \ref{Lambda} for three different types of binaries and for different values of $z_{\sup}$. Note that integrating the BNS, $z_{\sup}=6$ curve over the Advanced LIGO frequency band (roughly 10-200 Hz) yields $\Lambda \sim 10$ - in other words, in any 0.1 sec long time-segment (corresponding to the lowest observable frequency of 10 Hz) there will be on average 10 binary neutron star systems emitting in the 10-200 Hz band. The duty cycle is somewhat lower for the BBH and BHNS cases.

\begin{figure}[h]
\includegraphics[width=3in]{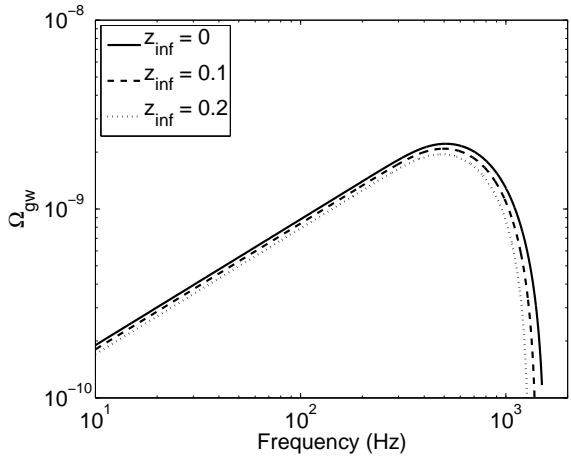}
\caption{Gravitational wave spectrum $\Omega_{gw}(f)$ computed for the BNS case with $M_c = 1.22 M_{\odot}$, $\lambda = 3\times 10^{-5} M_{\odot}^{-1}$, star formation rate from \cite{hopkins}, and $P(t_d) \sim t_d^{-1}$ with $t_{\rm min} = 20$ Myr. The effect of removing the nearest
binaries (with redshifts $z<0.1$ or $z<0.2$) is very small.}
\label{omegavsf}
\end{figure}

Figure \ref{Lambda} also shows that most of the contributing binaries reside at redshifts $z>0.1$. The nearest binaries are expected to produce loud chirp-like signals that could be individually detected by the upcoming detectors. However, such loud transients are typically explicitly excluded from the searches for GWB \cite{allenromano,S5stoch,S3stoch,S4stoch}. We have verified that the nearest (and loudest) binaries contribute little to $\Omega_{gw}$: Figure \ref{omegavsf} shows the gravitational-wave spectrum $\Omega_{gw}(f)$ computed for the BNS case with $M_c = 1.22 M_{\odot}$, $\lambda = 3\times 10^{-5} M_{\odot}^{-1}$, star formation rate from \cite{hopkins}, and $P(t_d) \sim t_d^{-1}$ with $t_{\rm min} = 20$ Myr. Excluding the nearest binaries (e.g. those with redshifts $z<0.1$) leads to a small ($<2\times$) reduction in the spectrum amplitude. We have further verified this conclusion with explicit Monte Carlo simulations for the case of Advanced LIGO collocated detector pair. Note that a similar Monte Carlo simulation was performed in the context of the Einstein Telescope \cite{ETMDC}.

\section{3. Results}
We performed a systematic study of the GWB due to binary coalescences, described in Section 2. In particular, we performed a scan of the parameter space spanned by the parameters $\lambda$ and $M_c$ for each of the BNS, BBH, and BHNS cases. We compare the model predictions for each point in the parameter space with sensitivities of current and future gravitational wave detectors, as well as with the estimates of the local coalescence rates presented in \cite{LIGOrates}. We also investigate the importance of the choice of star formation rate and of the choice of the probability distribution for the time delay between the formation of a binary and its coalescence.

\begin{figure*}[hbtp]
\centering
$\begin{array}{cc}
\includegraphics[width=0.45\textwidth,height=0.27\textheight]{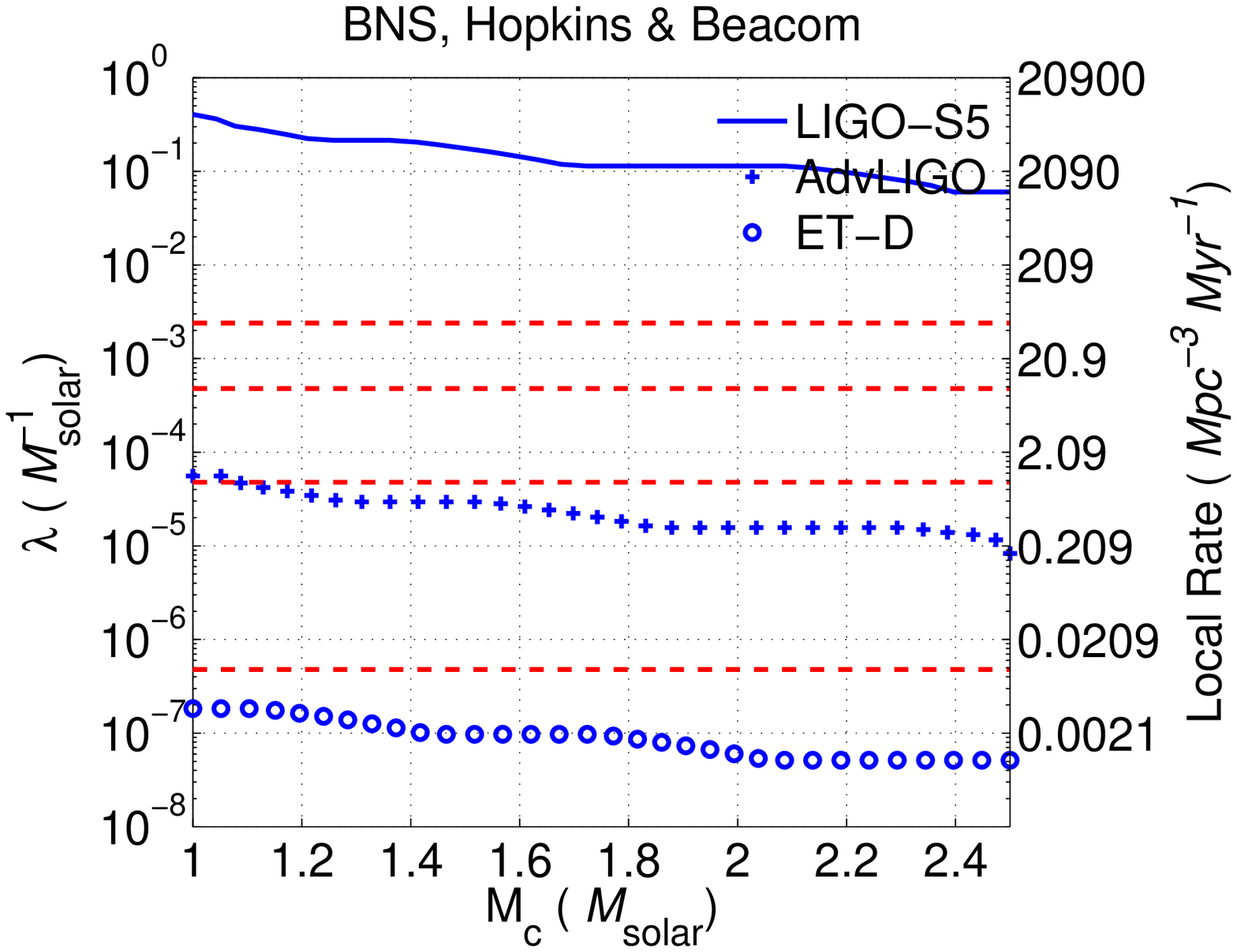} &
\hspace{1cm}
\includegraphics[width=0.45\textwidth,height=0.27\textheight]{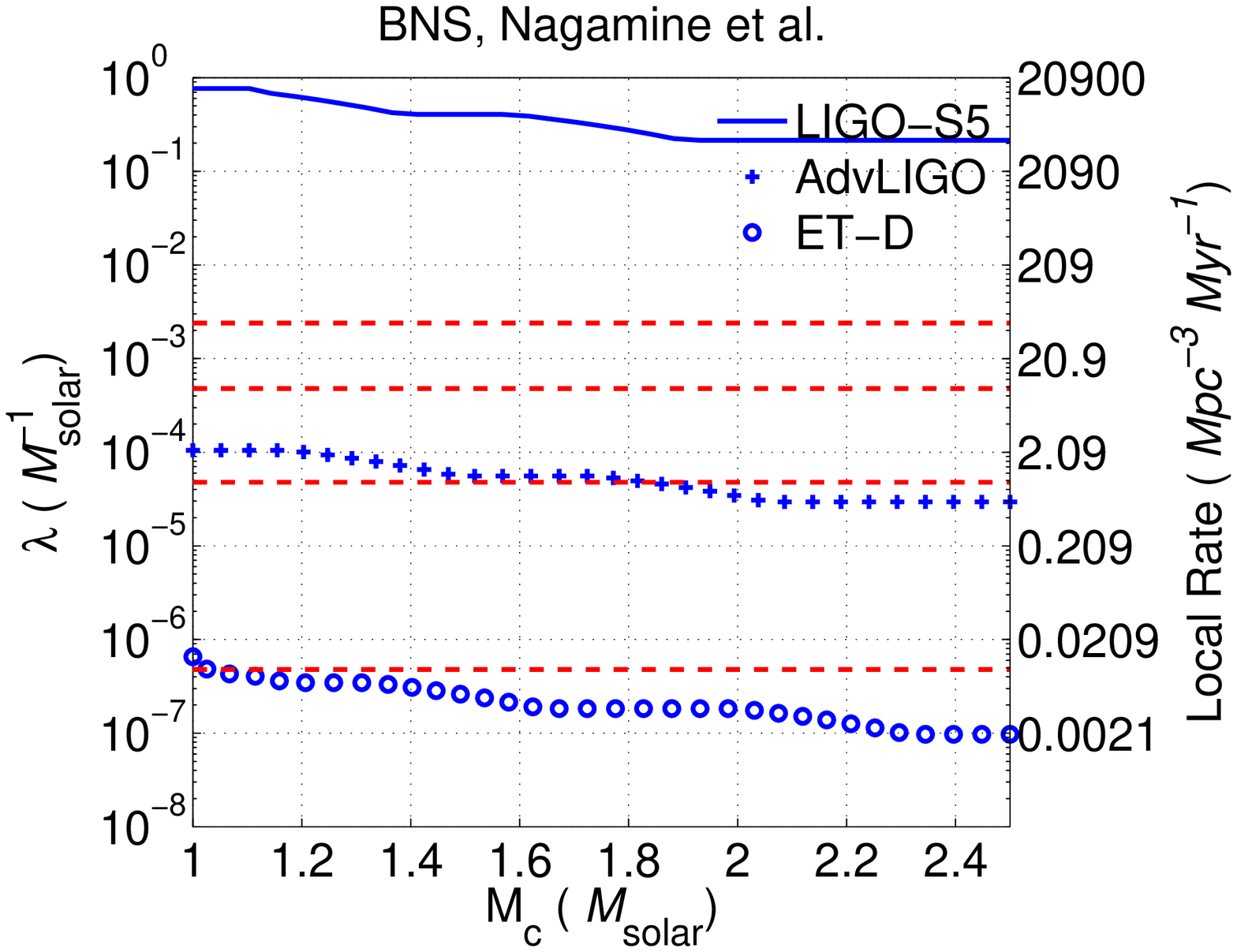}  \\
\includegraphics[width=0.45\textwidth,height=0.27\textheight]{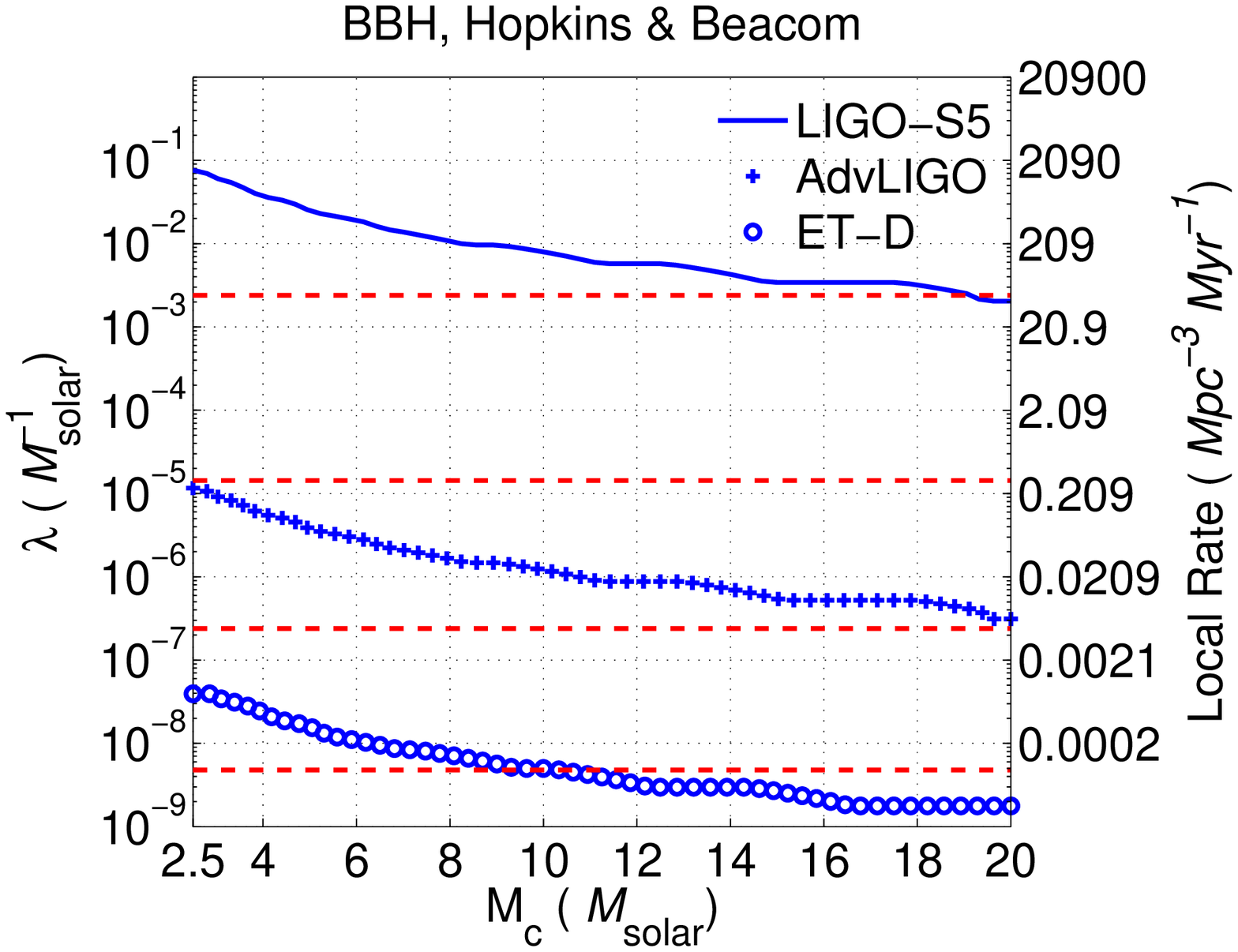} &
\hspace{1cm}
\includegraphics[width=0.45\textwidth,height=0.27\textheight]{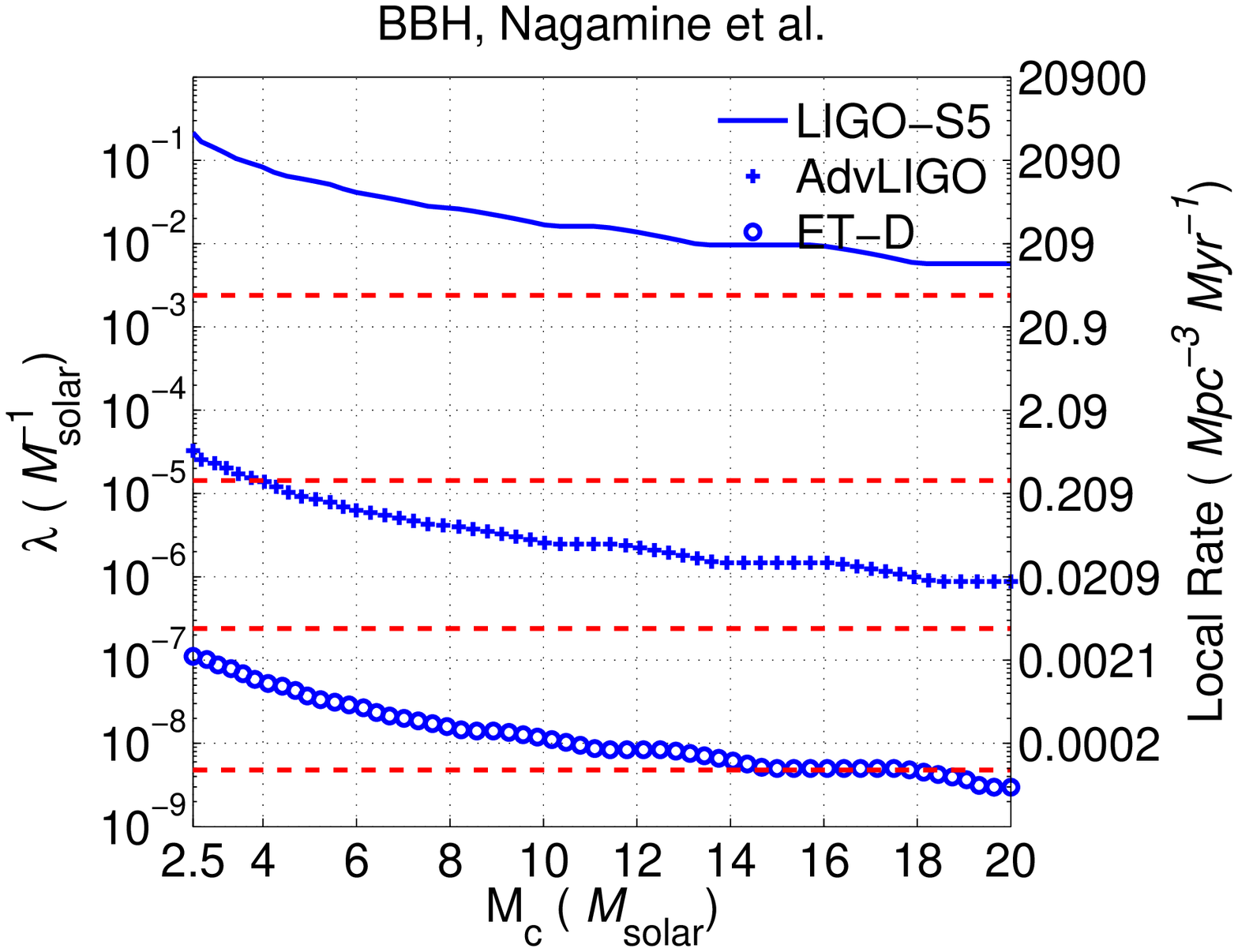} \\
\includegraphics[width=0.45\textwidth,height=0.27\textheight]{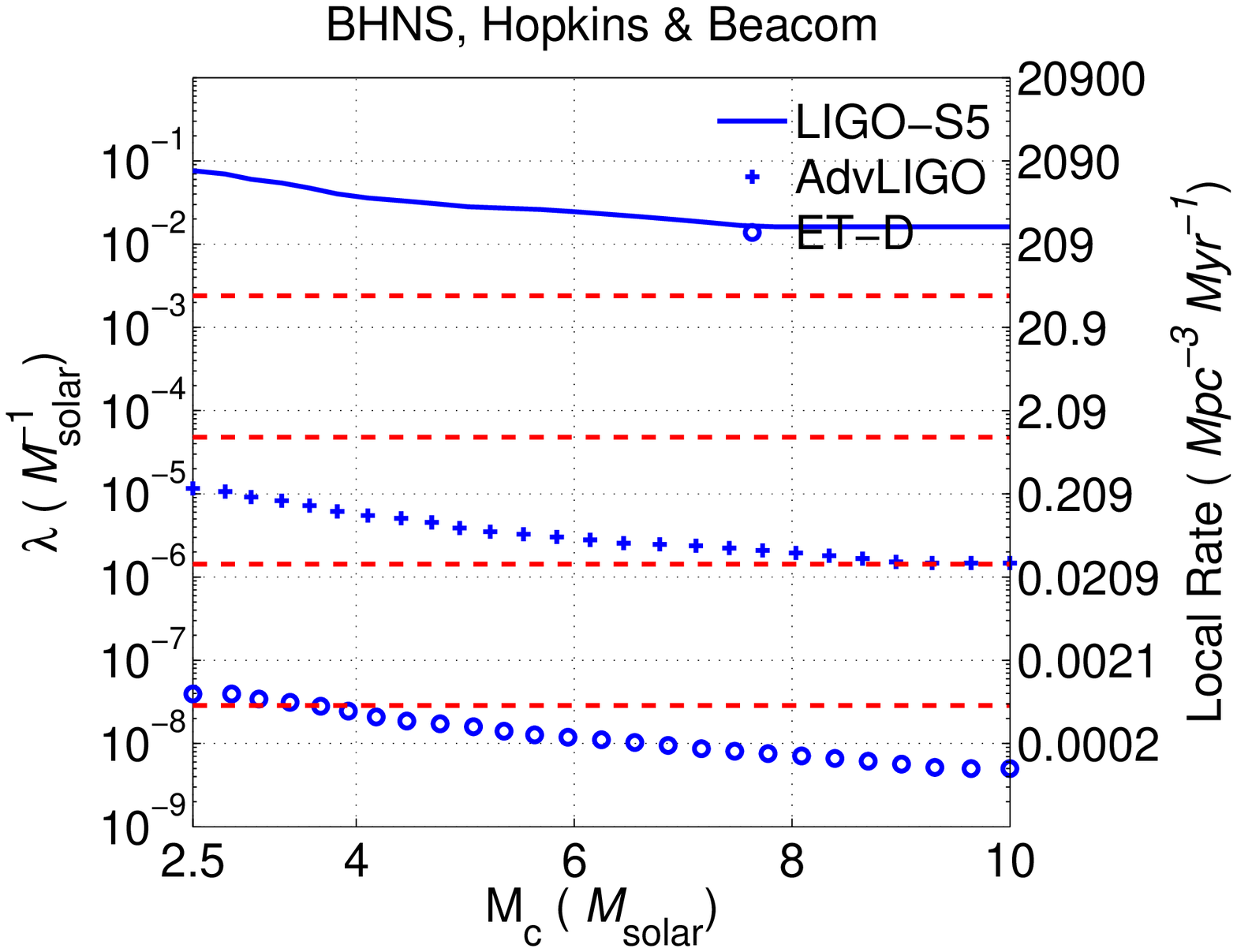} &
\hspace{1cm}
\includegraphics[width=0.45\textwidth,height=0.27\textheight]{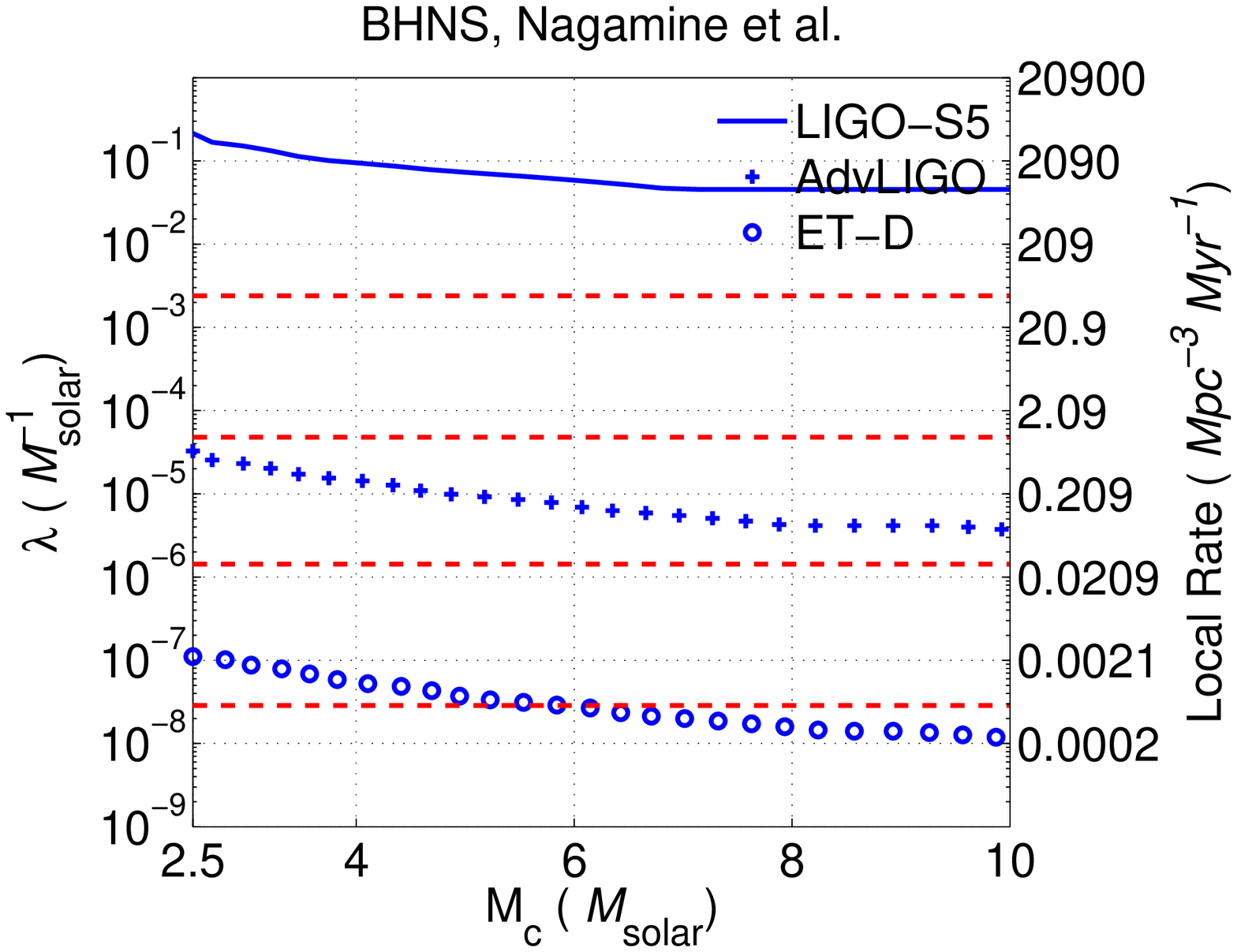} \\
\end{array}$
\caption{Accessibility of binary coalescence GWB to current and future gravitational wave detectors. The two columns correspond to two estimates of the star formation rate: Hopkins \& Beacom \cite{hopkins} (left) and Nagamine et al \cite{nagamine} (right). The three rows correspond to BNS, BBH, and BHNS respectively, top to bottom. For each plot we show the $\lambda-M_c$ plane: the region of the parameter space excluded by the S5 LIGO result \cite{S5stoch}, and the expected sensitivities of the Advanced LIGO collocated detector pair (assuming 1 year of exposure \cite{aLIGO,aLIGO2}), and of the Einstein Telescope (assuming two collocated detectors with ET-D sensitivity and one year of exposure \cite{ET,ET2}). These regions are to be compared with the expected local coalescence rates shown as horizontal dashed lines: top-to-bottom they correspond to maximal, optimistic, realistic, and pessimistic estimates presented in \cite{LIGOrates}.}
\label{scanplots}
\end{figure*}

To start with, we investigate the redshift dependence of the integrand in Equation \ref{omega} for different choices of the star formation rate. More specifically, the integrand can be written as:
\begin{equation}
I(z)  = \frac{R_V(z)}{E(\Omega_{\rm M}, \Omega_{\Lambda},z) (1+z)^{1/3}}.
\end{equation}
\begin{figure}[h]
\includegraphics[width=3in]{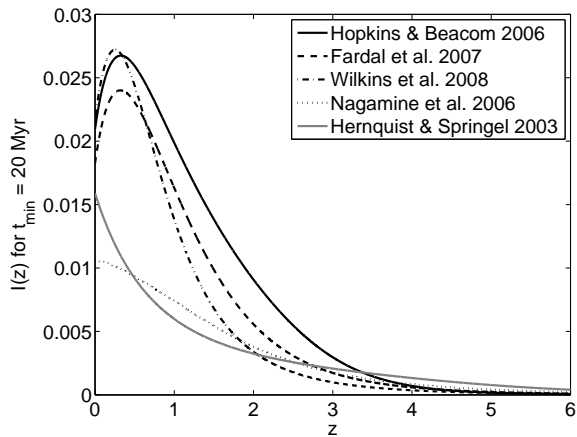}
\caption{$I(z)$ for different star formation rates, assuming $\alpha=-1$ and $t_{\min} = 20$ Myr.}
\label{sfr}
\end{figure}
Figure \ref{sfr} shows $I(z)$ for five different choices of the star formation rate. Since there are non-negligible differences between these five estimates of the star formation rate, we will present the results for the two extreme cases, namely the Hopkins \& Beacom \cite{hopkins} and Nagamine et al \cite{nagamine}. In particular, we scan the parameter space in the $\lambda-M_c$ plane: for each point in this parameter space, we compute $\Omega_{gw}(f)$ and we compare it to the most recent upper limit from LIGO \cite{S5stoch} and to the projected sensitivities for the second-generation (Advanced LIGO) and the third-generation (Einstein Telescope) gravitational-wave detectors. Figure \ref{scanplots} shows the results of the scan of the $\lambda-M_c$ plane for the two estimates of the star formation rate and for the three binary coalescence cases: BNS, BBH, and BHNS. For each of these cases, we observe that the latest GWB upper limit obtained using the LIGO data \cite{S5stoch} excludes the largest values of $\lambda$, which are larger than the maximal expected local coalescence rates \cite{LIGOrates}. However, the values of $\lambda$ corresponding to the optimistic and realistic (in the case of BNS and BHNS) local coalescence rates will be accessible to the second generation network of gravitational wave detectors (assuming standard Advanced LIGO expected strain sensitivity \cite{aLIGO,aLIGO2} for two collocated detectors with one year of exposure). The values of $\lambda$ corresponding to the pessimistic local coalescence rates will be accessible to the third-generation gravitational-wave detector network (assuming ET-D strain sensitivity curve \cite{ET,ET2} for two collocated detectors and one year of exposure). In fact, the binary coalescence GWB will be a foreground masking the GWB background due to early-Universe sources (inflationary models \cite{starob,barkana}, or phase transitions models \cite{phasetrans}) which may be one of the targets of the third generation detectors. Hence, to detect the cosmological GWB it will be necessary to identify and subtract all of the inspiral signals from all binaries in the frequency band of the third generation detectors. This is a daunting task, but appears to be plausible as demonstrated in \cite{CH_BBO} for the framework of the Big Bang Observer satellite-based detector \cite{BBO}. These conclusions hold for all of the three cases (BNS, BBH, and BHNS) and are rather weakly dependent on the choice of the star formation rate (we have verified that using the remaining three estimates of the star formation rate
\cite{fardal,wilkins,springel} yields results roughly between those shown here for the Hopkins \& Beacom \cite{hopkins} and Nagamine et al \cite{nagamine}).

\begin{figure*}[hbtp]
$\begin{array}{cc}
\includegraphics[width=0.45\textwidth,height=0.27\textheight]{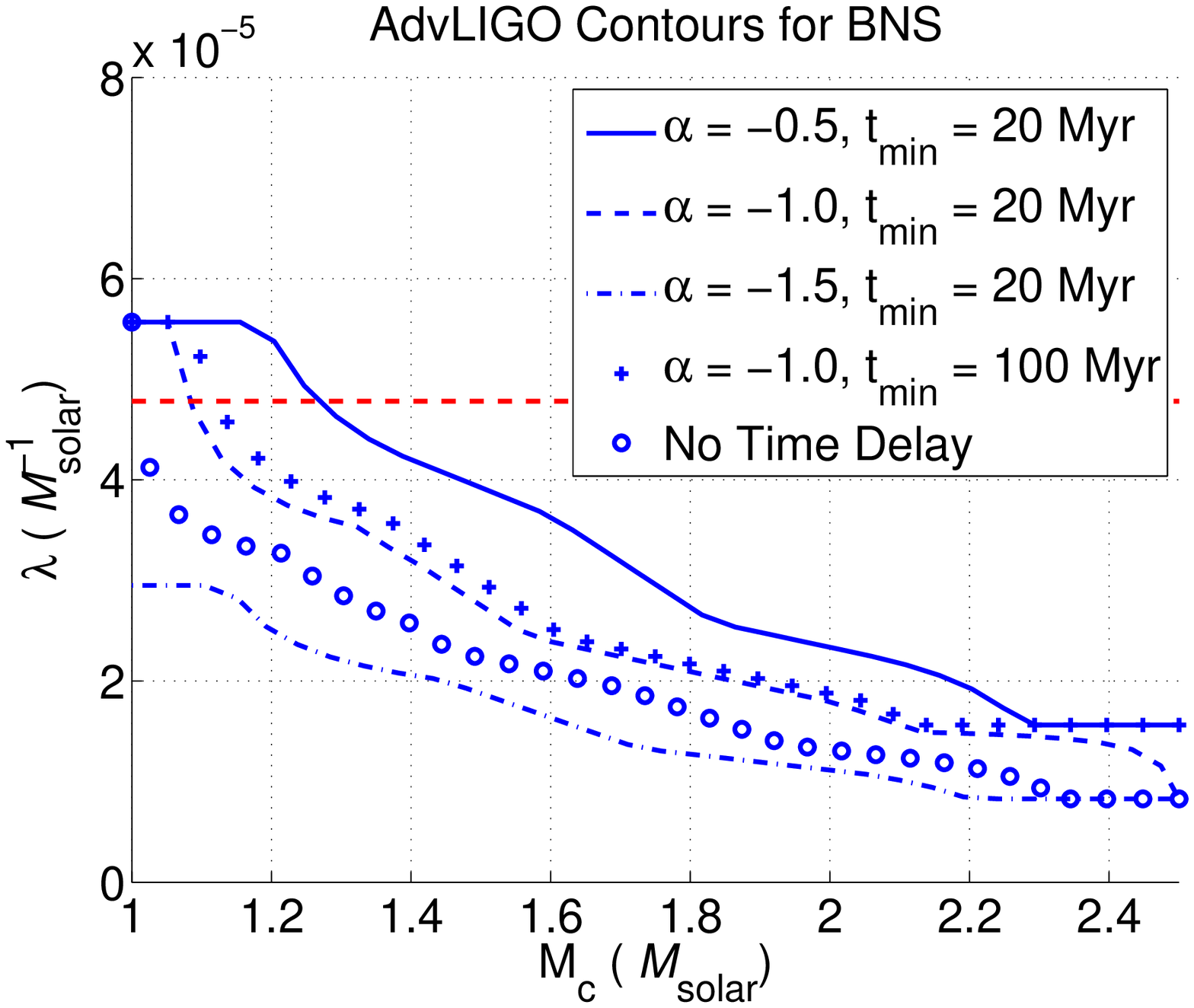} &
\hspace{1cm}
\includegraphics[width=0.45\textwidth,height=0.27\textheight]{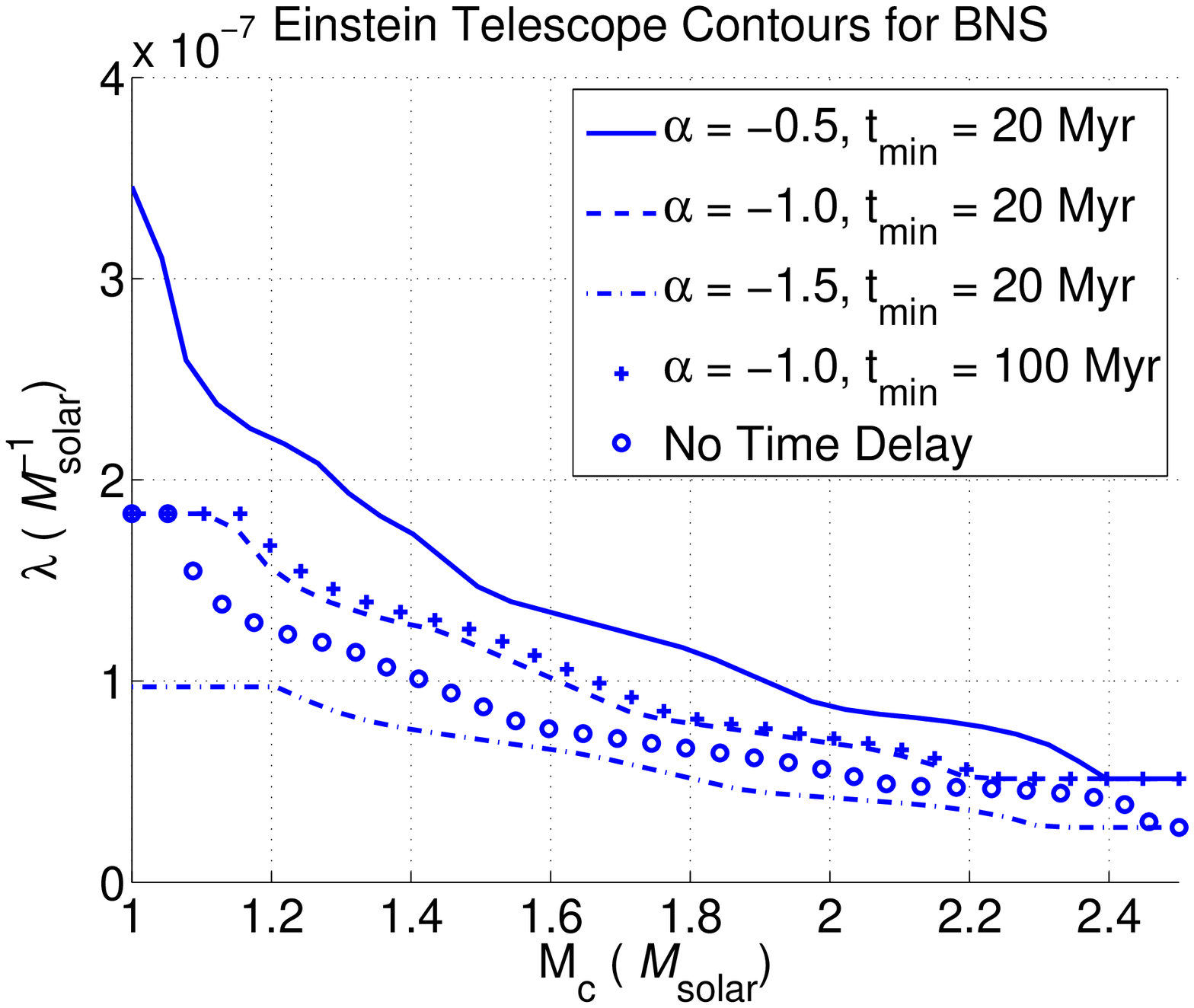} \\
\includegraphics[width=0.45\textwidth,height=0.27\textheight]{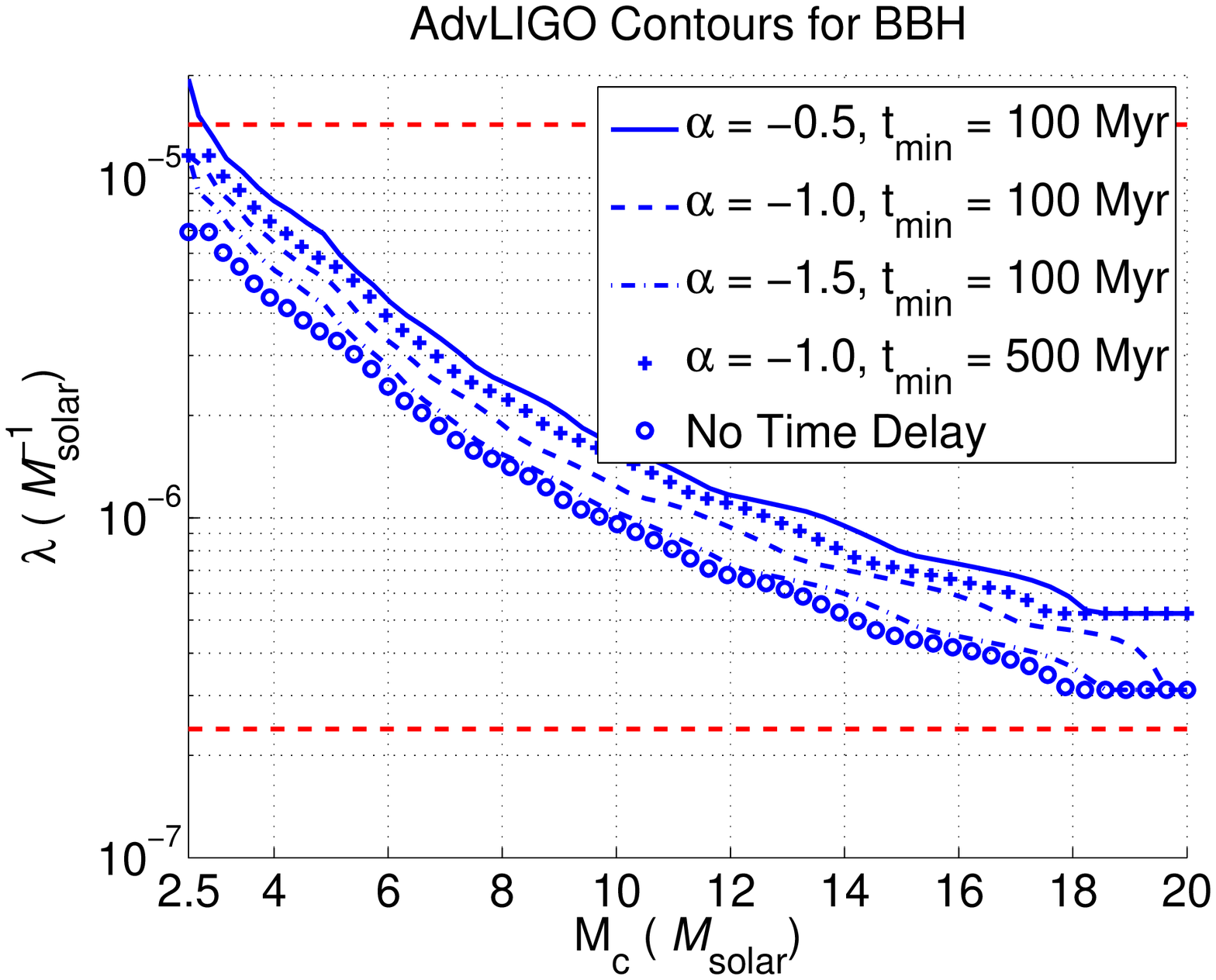} &
\hspace{1cm}
\includegraphics[width=0.45\textwidth,height=0.27\textheight]{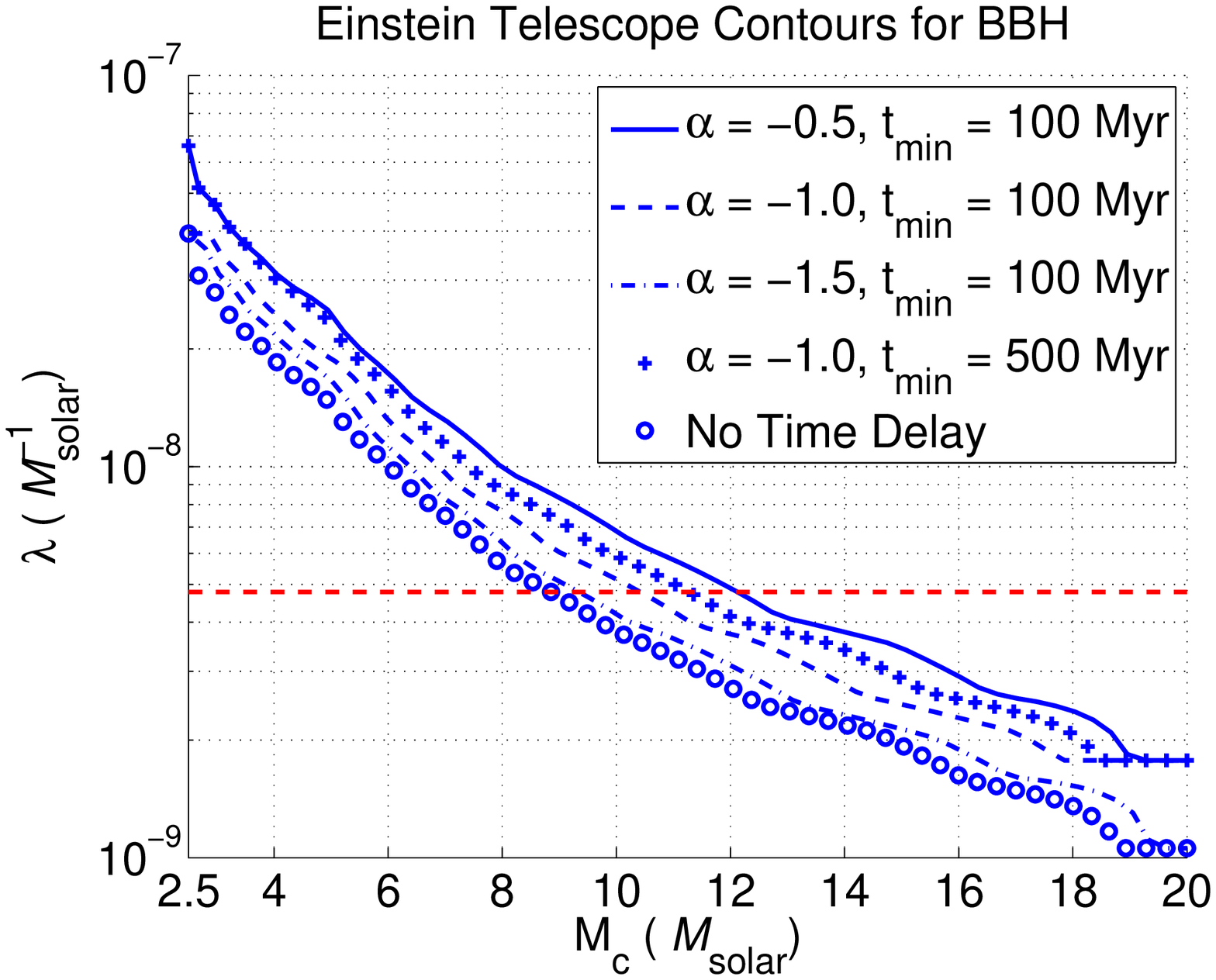} \\
\end{array}$
\caption{Effect of different $P(t_d)$ choices on the contours for Advanced LIGO (first column) and ET (second column) for the BNS (first row) and BBH (second row). Dashed horizonal lines correspond to realistic estimate (upper-left), optimistic and realistic estimates (lower-left), and pessimistic estimate (lower-right). In the upper-right plot, all contours are below the pessimistic estimate.}
\label{fig_variations}
\end{figure*}

We also investigated the effect of different choices of the probability distribution $P(t_d)$. As noted above, the population synthesis suggest $P(t_d) \sim t_d^{\alpha}$ for $t_d>t_{\min}$, where $t_{\min}$ is the minimum delay time for a massive binary to evolve until coalescence. Since there is some uncertainty in the parameters $\alpha$ and $t_{\min}$, we will probe the range of values of these parameters discussed in the literature. In particular, we examine $\alpha = -0.5, -1, -1.5$, $t_{\min} = 20, 100$ Myr for BNS and $t_{\min} = 100, 500$ Myr for BBH, as well as the case when there is no time delay between the formation and coalescence of the binary. Figure \ref{fig_variations} shows the variation in the contours for the second and third generation detectors, for BNS and BBH models, for several of the $P(t_d)$ parametrizations. We observe that the contours are rather insensitive to $P(t_d)$, varying by at most a factor of 2 in the $\lambda$ parameter. Hence, the choice of $P(t_d)$ does not qualitatively affect the conclusions of this study.

Finally, we note that for the case of BBH a similar study was performed in \cite{BBHstoch} - they computed Advanced LIGO and ET-D contours in the $\lambda-M_c$ plane that were substantially higher in $\lambda$ ($\sim 20\times$ for Advanced LIGO). These differences largely come from the different assumptions in detector sensitivity and exposure. For Advanced LIGO they assumed a non-standard detector strain sensitivity (this is the dominant cause of discrepancy), non-collocated detectors, and 3 years of exposure - we assume the standard strain sensitivity, collocated detectors, and one year of exposure. For ET-D, they assumed the triangular detector configuration (leading to the factor of $3/8$ in overlap reduction), while we assumed L-shaped interferometers.
%
%

\section{4. Conclusions}
In this paper we computed the gravitational wave background due to coalescences of binary neutron stars, binary black holes, and black hole - neutron star binaries, following the approach of \cite{regman}. While such computations have been done in the past, in this study we performed a systematic scan of the parameter space, taking into account the possible variations in the result due to the choice of the star formation rate, and due to the choice of the distribution $P(t_d)$ of the delay time between the formation and coalescence of the binary. For each point in the parameter space, we compare the model prediction to the expected sensitivities of the second and third generation gravitational-wave detector networks. We find that models corresponding to the optimistic and realistic (in the case of BNS and BHNS models) local coalescence rates will be accessible to the second generation detector network. We also find that models corresponding to the pessimistic local coalescence rates will be accessible to the third-generation detector network. The binary coalescence GWB will, in fact, be a foreground for the third-generation detectors, and it will mask the GWB background due to early-Universe sources. Accessing the cosmological GWB with third generation detectors will therefore require identification and subtraction of all inspiral signals from all binaries in the relevant frequency band.

{\it Acknowledgments:}
The authors thank C. Belczinski, M. Dominik, and T. Bulik for informative discussions about the probability distribution for delay time, $P(t_d)$. CW and VM were supported in part by NSF grant PHY0758036.


\end{document}